\documentclass[%
 reprint,
superscriptaddress,
bibnotes,
amsmath,amssymb,
aps,
prx,
floatfix,
showkeys
]{revtex4-2}

\usepackage{graphicx}% 
\usepackage{subcaption}% This package breaks the justification of the captions in RevTex
\usepackage{adjustbox}
\usepackage{float}
\usepackage{dcolumn}% Align table columns on decimal point
\usepackage{bm}% bold math
\usepackage{xr}% reference to external document
\usepackage{orcidlink}
\usepackage[mathlines]{lineno}% Enable numbering of text and display math
\usepackage{color}
\usepackage{lmodern}
\usepackage{booktabs}
\usepackage[separate-uncertainty=true, multi-part-units=single]{siunitx}
\usepackage[OT1]{fontenc}
\usepackage{placeins}
\usepackage{ragged2e}
\usepackage{stfloats}
\usepackage{float}
\usepackage{ulem}

\usepackage{sansmath} % Enables turning on sans-serif math mode, and using other environments
\sansmath % Enable sans-serif math for rest of document
\usepackage{multirow}

\newcommand{\exciting}{{\usefont{T1}{lmtt}{b}{n}exciting}}

\DeclareSIUnit\angstrom{\text {Å}}
\DeclareSIUnit\bar{bar}
\makeatletter
\newcommand*{\addFileDependency}[1]{% argument=file name and extension
\typeout{(#1)}% latexmk will find this if $recorder=0
\@addtofilelist{#1}
\IfFileExists{#1}{}{\typeout{No file #1.}}
}\makeatother

\newcommand*{\myexternaldocument}[1]{%
\externaldocument{#1}%
\addFileDependency{#1.tex}%
\addFileDependency{#1.aux}%
}

\myexternaldocument{supporting_infomation}

\begin{document}

%\title{First-Principles Approach to Disentangling Electronic and Thermal Contributions in Transient Spectra of Semiconductors}

\title{First-principles approach to ultrafast pump-probe spectroscopy in solids}

\author{Lu Qiao\orcidlink{0009-0002-0344-1082}}
%\thanks{Both authors contributed equally to this work.}
\affiliation{Department of Physics and CSMB, Humboldt-Universit\"at zu Berlin, Zum Gro\ss en Windkanal 2, D-12489 Berlin, Germany}
\email{qiaolu@physik.hu-berlin.de}
\author{Ronaldo Rodrigues Pela\orcidlink{0000-0002-2413-7023}}
%\thanks{Both authors contributed equally to this work.}
\affiliation{Supercomputing Department, Zuse Institute Berlin (ZIB), Takustraße 7, 14195 Berlin, Germany}

\author{Claudia Draxl\orcidlink{0000-0003-3523-6657}}
\affiliation{Department of Physics and CSMB, Humboldt-Universit\"at zu Berlin, Zum Gro\ss en Windkanal 2, D-12489 Berlin, Germany}

\date{\today}

\begin{abstract}
Pump-probe spectroscopy is a powerful tool to study ultrafast exciton dynamics, revealing the underlying complex interactions on the electronic scale. Despite significant advances in experimental techniques, developing a comprehensive and rigorous theoretical framework for modeling and interpreting the transient response in photoexcited materials remains a challenge. Here, we present a first-principles approach to simulating pump-probe spectroscopy and disentangling the electronic and thermal contributions underlying exciton dynamics. We showcase our method to three materials, representative for different classes of solids: the transition-metal dichalcogenides WSe$_2$, the halide perovskite CsPbBr$_3$, and the transition-metal oxide TiO$_2$, showing remarkable agreement with experimental counterparts. We find that (i) photoinduced Coulomb screening is the primary electronic effect, responsible for a blue shift of exciton resonances, while (ii) Pauli blocking plays a minor role, and (iii) thermal lattice expansion leads to a red shift of the spectra. We further demonstrate how key parameters such as excitation density, pump photon energy, and pump polarization modulate the transient absorption spectra, offering direct control over the exciton-resonance energy. Our approach establishes a quantitative and predictive framework for interpreting pump-probe experiments, providing actionable insights for the design of energy-selective optoelectronic devices through exciton engineering.
\end{abstract}
\maketitle

%%%%%%%%%%%%%%%%%%%%%%%%%%%%%%%%%%%%%%%%%%%%%%%%%%%%%%%%%%%%%%%%%%%%
%%%%%%%%%%%%% Introduction of pump-probe spectroscopy  %%%%%%%%%%%%%
%%%%%%%%%%%%%%%%%%%%%%%%%%%%%%%%%%%%%%%%%%%%%%%%%%%%%%%%%%%%%%%%%%%%
Pump-probe spectroscopy is a state-of-the-art time-resolved technique for investigating non-equilibrium exciton dynamics and light-matter interactions \cite{ashoka2022}. In a typical pump-probe experiment, the material is optically excited by a pump pulse, and the resulting transient changes in transmission are measured with a time-delayed probe pulse, which encodes information about exciton dynamics and structural response on ultrafast timescales \cite{beeby2002}. Advances in laser technology enable this technique to access a broad range of timescales and energy windows, from attosecond to picosecond regimes \cite{lucchini2021,ashoka2022,barends2024}, and from the infrared and visible \cite{montanaro2020} to the extreme ultraviolet (XUV) \cite{rebholz2021} and X-ray domains \cite{geneaux2019}.

%%%%%%%%%%%%%%%%%%%%%%%%%%%%%%%%%%%%%%%%%%%%%%%%%%%%%%%%%%%%%%%%%%%%
%%%%%%%%%%%%%%%%% Theory for pump-probe spectroscopy  %%%%%%%%%%%%%%
%%%%%%%%%%%%%%%%%%%%%%%%%%%%%%%%%%%%%%%%%%%%%%%%%%%%%%%%%%%%%%%%%%%%
The growing interest in pump-probe experiments has driven a parallel effort to develop theoretical methods aimed at modeling transient phenomena. Density-matrix theory was among the first theoretical approaches developed to describe transient absorption (TA) spectra \cite{lindberg1988,pollard1990,yan1997,wolfseder1997}. However, these approaches rely on model Hamiltonians and thus lack first-principles accuracy. Beyond model-based approaches, real-time time-dependent density-functional theory (RT-TDDFT) is widely used as a first-principles method to simulate TA spectroscopy \cite{de2013,pemmaraju2020,moitra2023}. While efficient and applicable to large systems, RT-TDDFT within a mean-field framework lacks an explicit treatment of many-body interactions. In contrast, many-body wavefunction approaches, such as the time-dependent configuration interaction singles approach \cite{pabst2012} and coupled cluster theory \cite{skeidsvoll2020}, offer a high-accuracy description of electron correlation. Yet, they are computationally quite involved and thus limited to small atomic or molecular systems. Many-body perturbation theory, based on a Green-function formalism, can accurately capture many-body interactions, and one can simulate equilibrium absorption spectra of periodic systems with the Bethe–Salpeter equation (BSE). Nevertheless, the effect of photoexcited carriers is not included in standard implementations. In recent work, this formalism has been generalized to the non-equilibrium regime and applied to simulate the TA spectra of a simple four-band model system (two valence and two conduction states) \cite{perfetto2015}, and to the transient reflectivity spectrum of Silicon \cite{sangalli2016}. Ultimately, a comprehensive study of TA spectra across diverse materials and timescales is still missing. 

%%%%%%%%%%%%%%%%%%%%%%%%%%%%%%%%%%%%%%%%%%%%%%%%%%%%%%%%%%%%%%%%%%%%
%%%%%%%%%%%%%%%%%%%%% A brief summary for this work   %%%%%%%%%%%%%%
%%%%%%%%%%%%%%%%%%%%%%%%%%%%%%%%%%%%%%%%%%%%%%%%%%%%%%%%%%%%%%%%%%%%
To address this challenge, we develop a first-principles approach that integrates constrained density-functional theory (cDFT), RT-TDDFT, and a non-equilibrium extension of the BSE to simulate TA spectroscopy. We apply our method to three prototypical materials: tungsten diselenide (WSe$_2$), cesium lead bromide (CsPbBr$_3$), and anatase titanium dioxide (TiO$_2$). These compounds are representative of 2D transition metal dichalcogenides (TMDCs), halide perovskites, and transition-metal oxides, respectively, and are widely used in applications such as photodetection~\cite{peng2021}, photocatalysis~\cite{zheng2016,yu2017}, and solar energy conversion~\cite{lewerenz1980,ullah2021}. By using our newly developed method, we not only obtain the TA spectra in good agreement with experimental results, but can disentangle the spectral features into electronic and thermal contributions. It provides a powerful tool to interpret pump–probe experiments and gain insights into exciton dynamics, offering guidelines for exciton engineering in high-performance optoelectronic devices.

\noindent\textbf{First-principles formalism to simulate TA spectra.} To simulate TA spectra, we extend the standard BSE to a non-equilibrium formalism by explicitly incorporating excited-state carrier occupations derived from cDFT or RT-TDDFT. These two approaches provide frameworks for describing thermal and non-thermal carrier distributions on ultrafast timescales, respectively. Details of the formalism are provided in the Supplementary Note \ref{secSI:theory_background}, and the workflow is summarized as follows: \\

\textbf{(i) Thermal carrier distribution on the picosecond timescale.}
At picosecond time delays following the optical pumping, the carrier populations are assumed to be thermalized at the carrier temperature. This is modeled within cDFT. Excited-state occupations follow the Fermi-Dirac distribution, with electrons and holes filling the conduction band minimum (CBM) and valence band maximum (VBM), respectively. \\

\textbf{(ii) Non-thermal carrier distribution on the femtosecond timescale.} The time-evolution of carrier distributions at femtosecond delays following optical pumping is simulated using RT-TDDFT. We employ the velocity-gauge with the time-dependent vector potential $\mathbf{A}(t)$ to model coherent light-matter interaction. The orbitals evolve according to the time-dependent Kohn-Sham equations, and the excited carrier distribution is obtained by projecting the orbitals onto the ground-state wavefunctions. This yields a $\mathbf{k}$- and state-resolved non-thermal carrier distribution that captures the excitation pathway and energy-momentum redistribution driven by the optical field.\\

\textbf{(iii) Electronic contribution to TA spectra.} 
The thermal and non-thermal occupations are subsequently used to solve the non-equilibrium BSE and calculate the frequency-dependent macroscopic dielectric function at a time delay $t$ after the optical pump:
\begin{equation}
   \mathrm{\Delta Im}\varepsilon(\omega, t) = \mathrm{Im}\varepsilon^{\mathrm{neq}}(\omega, t) - \mathrm{Im}\varepsilon^{\mathrm{eq}}(\omega),
\end{equation}
where $\mathrm{Im}\varepsilon^{\mathrm{eq}}(\omega)$ and $\mathrm{Im}\varepsilon^{\mathrm{neq}}(\omega)$ represent the equilibrium and non-equilibrium cases. This electronic TA spectrum includes two separable many-body effects: (i) Pauli blocking and (ii) Coulomb screening. Pauli blocking originates from the exclusion principle, which prevents excitations to occupied states. Coulomb screening refers to the reduction of Coulomb interaction between electrons and holes due to the screening by excited charge carriers. Pauli blocking and Coulomb screening are disentangled in our approach by either incorporating excited-state occupations to evaluate the dipole moment matrix or the screened Coulomb interaction, as detailed in equations \ref{eq:CoulombScreening} and \ref{eq: PauliBlocking} in the Supplementary Note \ref{secSI:theory_background}. \\

\textbf{(IV) Thermal contribution to TA spectra.} On the picosecond timescale, incoherent electron-ion collisions drive thermal equilibration between the electronic and lattice subsystems, yielding an almost uniform temperature throughout the solid \cite{bauerhenne2024}. Here, we simulate the pump-induced local heating through a homogeneous lattice expansion at this respective temperature. The thermal contribution to the TA spectrum, denoted as $\mathrm{\Delta Im}\varepsilon(\omega,T)$, is defined as the difference between the high-temperature spectrum $\mathrm{Im}\varepsilon(\omega,T)$ and the low-temperature reference $\mathrm{Im}\varepsilon^{eq}(\omega)$:
\begin{equation}
   \mathrm{\Delta Im}\varepsilon(\omega,T) = \mathrm{Im}\varepsilon(\omega,T) - \mathrm{Im}\varepsilon^{eq}(\omega)
\end{equation}
All these developments have been implemented in the full-potential code \exciting\thinspace\cite{gulans2014}, thereby extending its excited-state modules for RT-TDDFT and the BSE. Within the LAPW+lo all-electron framework, both core and valence electrons are treated on the same footing, which enables a consistent description of optical and core excitations. The approach presented here was successfully demonstrated by computing the electronic contributions to the transient absorption spectrum at the Zn K-edge of ZnO, obtaining excellent agreement with experiment \cite{Lu2025}. Building on this, we here discuss the methodology in more detail, advance the formalism to also account for thermal effects, probe different core edges, and perform a detailed analysis of results for different material prototypes.\\

\noindent\textbf{Transient absorption at the Se M edge in \texorpdfstring{WSe$_\mathsf{2}$}{WSe2}.}

\begin{figure*}
    \centering
    \includegraphics[width=0.8\linewidth]{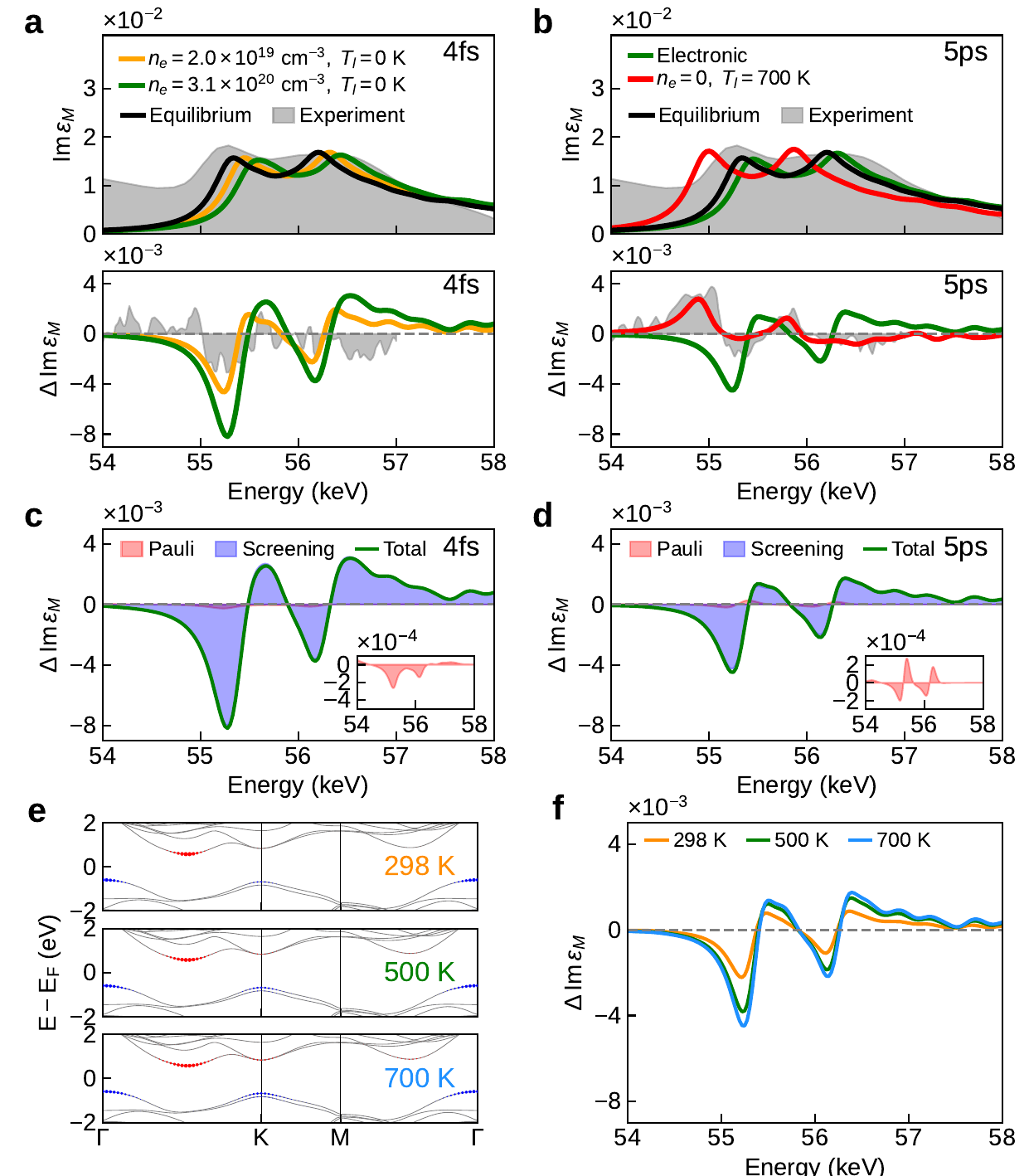}
    \caption{\textbf{Electronic and thermal contribution to TA spectra at the Se M$_\mathbf{\mathsf{4,5}}$-edges of WSe$_\mathsf{2}$.} (a) Calculated absorption (top) and TA (bottom) spectra on the femtosecond timescale, showing electronic contributions at different excitation densities. (b) Electronic (green) and thermal (red) contributions to the spectra on the picosecond timescale. Here, $n_{e}$ represents the excitation density, and $T_{l}$ the lattice temperature. Calculated equilibrium spectra (black) and experimental data (gray, from Ref.~\cite{Oh_2023}) are shown for comparison. (c, d) Decomposition of the electronic contributions into Coulomb screening (blue) and Pauli blocking (red) on the (c) femtosecond and (d) picosecond timescales. Insets: Zoom into the Pauli blocking effect. (e) Thermalized Fermi-Dirac distributions of electrons (red) and holes (blue) at different electronic temperatures for a fixed excitation density. (f) Corresponding TA spectra computed using the thermalized carrier distributions shown in (e).}
    \label{fig:WSe2}
\end{figure*}

Transition metal dichalcogenides are layered materials whose two-dimensional monolayers are held together by van der Waals forces. This enables easy stacking and tuning their properties \cite{ji2017thickness, niu2018thickness,Kolobov2016}. TMDCs exhibit strong photoluminescence, pronounced excitonic effects, and spin-control capabilities with applications in valleytronics \cite{manzeli20172d,xu2014spin}. Their efficient light absorption also supports applications in photovoltaics \cite{jariwala2017van-363}, photocatalysis \cite{peng2017roles}, and photodetection \cite{wu2021synthesis-833}.
% WSe2
Among the TMDCs, WSe$_2$ stands out due to various advantageous characteristics. Notably, selenides like WSe$_2$ tend to be more stable and resistant to oxidation in a humid atmosphere than their sulfide counterparts \cite{liu2013role}. Furthermore, its inherent high carrier mobility and ambipolar characteristics \cite{wang2018ambipolar} make WSe$_2$ an ideal candidate for efficient n- and p-channel field-effect transistors (FETs) \cite{chuang2014high}. These properties also enable the fabrication of light-emitting diodes with remarkable electroluminescence \cite{ross2014electrically}.

% Challenges in XTA analysis
While much of recent research on TMDCs has been focused on tuning their properties by varying the number of layers and exploring spin-, valley-, and exciton-related phenomena, ultrafast spectroscopy remains a critically important yet emerging field of investigation \cite{buades2021attosecond,britz2021carrierspecific,Schumacher_2023,pogna2016photoinduced-163}. Such studies are, nevertheless, indispensable for understanding the relaxation process of excited carriers and how the material's properties evolve under excitation. This knowledge can help advance next-generation optoelectronics based on TMDCs \cite{britz2021carrierspecific, ji2017thickness,duong2019modulating}. Here, we perform detailed first-principles calculations on pump-probe spectroscopy of WSe$_2$, taking the experimental setup described in \cite{Oh_2023} as reference.

In a first step, we discuss the extreme ultraviolet (XUV) absorption spectra at the Se M$_\mathsf{4,5}$ edges in equilibrium. As observed in the top panels of Figs. \ref{fig:WSe2}a-b, the calculated spectra (black line) agree well with the measured counterpart (shaded gray area), reproducing the two-peak structure and the relative peak positions. The additional feature observed experimentally at around \SI{52}{\electronvolt} originates from transitions involving the W O$_\mathsf{2}$ edge, which is not considered here. We emphasize the importance of many-body effects for achieving this level of agreement, since the independent-particle approximation (IPA) cannot even qualitatively describe the peak structure observed in the experimental spectra (Fig. \ref{figSI:xas_WSe2_BSE_vs_IPA}). This contradicts earlier studies that claimed IPA to yield good qualitative agreement \cite{Oh_2023,Schumacher_2023}.

The top panels of Figs. \ref{fig:WSe2}a-b also show the impact of photoexcited carriers on the absprption spectra: A blue shift is observed, which increases with excitation density (orange and green lines). For comparison, the effect of lattice expansion, corresponding to a temperature increase to 700 K, is displayed (red line). While excitonic effects induce a blueshift in the absorption spectra, thermal effects cause a redshift. The changes relative to the equilibrium spectra, shown in the bottom panels of Figs. \ref{fig:WSe2}a-b, indicate that the lattice expansion is the dominant effect on the picosecond timescale. This conclusion is supported by the excellent agreement with experiment. Conversely, photoexcited carriers are the prevailing factor on the femtosecond timescale: As can be seen in the bottom panel of Fig. \ref{fig:WSe2}b, the calculation for $2.0\times 10^{19}$ cm$^{-3}$ (orange line) reproduces experiment satisfactorily well (the shaded area). Finally, we observe that the TA spectra show nonlinear behavior as doubling the excitation density does not result in a twofold increase in intensity as seen in the bottom panel of Fig. \ref{fig:WSe2}a. 

Our methodology also enables us to disentangle the contributions to the TA spectra into Pauli blocking and Coulomb screening by photoexcited carriers, as illustrated in Figs. \ref{fig:WSe2}c-d,. Pauli blocking (shaded red) is negligible on both the femtosecond and picosecond timescales. Conversely, changes in Coulomb screening (blue-colored areas) caused by the photoexcited electrons and holes primarily account for all features observed in the TA spectra.

Finally, it is also interesting to examine how the TA spectra depend on the carrier distribution in k-space. For a fixed carrier density of $1.0\times 10^{20}$ cm$^{-3}$, Fig. \ref{fig:WSe2}e displays how the excited electron (red circles) and hole (blue circles) population evolve as the electronic temperature increases from 298 K to 500 K and 700 K. Lattice relaxation is not included to mimic different scenarios occurring on a picosecond timescale after pump excitation. At lower temperatures, electrons and holes are more localized around the CBM and VBM, respectively. As the temperature increases, the carrier distribution becomes more delocalized. This effect is especially visible at the K point.

The impact of carrier distribution is depicted in Fig. \ref{fig:WSe2}f. Increasing the electronic temperature results in a more scattered distribution of excited electrons and holes across k-points, leading to enhanced screening. This, in turn, amplifies the TA response, i.e., positive and negative features become more pronounced at higher temperatures. Notably, this effect tends to saturate around 500 K.\\

\noindent\textbf{Transient absorption at the Br K-edge in \texorpdfstring{CsPbBr$_3$}{CsPbBr3}.}
%%%%%%%%%%%%%%%%%%%%%%%%%%%%%%%%%%%%%%%%%%%%%%%%%%%%%%%%%%%%
%%%%%%%%%%%%%%%%% Introduction of CsPbBr3  %%%%%%%%%%%%%%%%%
%%%%%%%%%%%%%%%%%%%%%%%%%%%%%%%%%%%%%%%%%%%%%%%%%%%%%%%%%%%%
Halide perovskites have garnered significant interest due to their remarkable improvement in solar-cell energy-conversion efficiency over the past decade \cite{nrel2024}. Among these, CsPbBr$_3$ stands out for its exceptional stability \cite{duan2021}. Although optical TA spectroscopy has been widely used to study carrier dynamics in CsPbBr$_3$ \cite{liu2022,mondal2017}, the use of X-ray TA spectroscopy, which provides simultaneous sensitivity to electronic and lattice degrees of freedom, remains limited. Moreover, a comprehensive theoretical interpretation of X-ray TA spectra in perovskites is fully lacking. Here, we report a detailed investigation of photoexcited CsPbBr$_3$, using our first-principles method to disentangle electronic and thermal contributions to the transient spectral response.

\begin{figure*}
    \centering
    \includegraphics[width=0.8\linewidth]{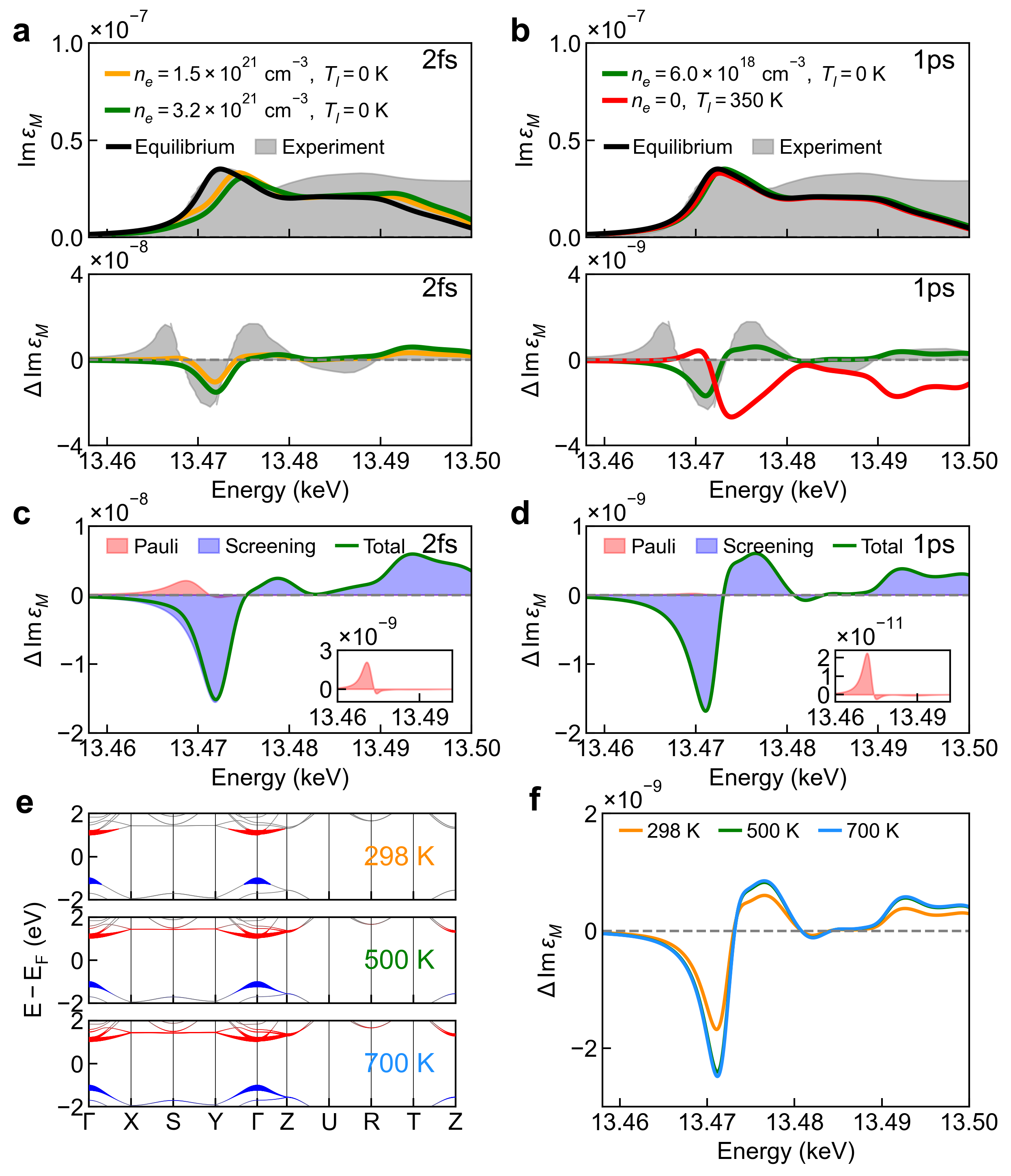}    
    \caption{\textbf{Electronic and thermal contributions to the TA spectra at the Br K-edge of CsPbBr$_3$.} Same as  Fig.~\ref{fig:WSe2}, but for the case of the Br K-edge of CsPbBr$_3$.}
    \label{fig:CsPbBr3}
\end{figure*}

%%%%%%%%%%%%%%%%%%%%%%%%%%%%%%%%%%%%%%%%%%%%%%%%%%%%%%%%%%%%
%%%%%%%%%%%%%% absorption spectra at ps and fs timescale %%%%%%%%%%
%%%%%%%%%%%%%%%%%%%%%%%%%%%%%%%%%%%%%%%%%%%%%%%%%%%%%%%%%%%%
Before examining the effects of electronic (photoexcited carriers) and thermal (lattice expansion) contributions to the TA spectra, we first compare in Figs.~\ref{fig:CsPbBr3}a-b (top panels) the calculated equilibrium absorption spectra at the Br K-edge to experiment (gray-shaded regions) \cite{cannelli2021}. The simulations show good agreement with the experiment. Upon incorporating the effect of photoexcited carriers, the spectra (orange and green lines) display a blue shift that increases with excitation density. At femtosecond time delays, this blue shift is particularly pronounced. The electronic TA amplitude at a time delay of \SI{2}{\femto\second} is approximately an order of magnitude larger than that observed on the picosecond timescale, owing to the higher excitation density shortly after the pump pulse, when carrier relaxation and recombination are still negligible. (top panel of Figure~\ref{fig:CsPbBr3}a). The electronic TA signal successfully reproduces most experimental features, except for the initial positive peak around $\sim$\SI{13.466}{\kilo\electronvolt}, which likely involves thermal expansion effects as studied here. Thermal effects are only analyzed on the picosecond timescale, as carrier cooling occurs on sub-picosecond timescales and thermalization is expected to dominate at longer delay times. Lattice expansion at 350 K (red lines) leads to a red shift in the TA spectra, arising from the reduced separation between the Br 1s core level and the conduction states (Figure~\ref{fig:CsPbBr3}b). 

For both the picosecond and femtosecond electronic TA spectra, the dominant effect arises from Coulomb screening (blue-shaded areas in Figs.~\ref{fig:CsPbBr3}c-d). Pauli blocking (red-shaded areas) is negligible on the picosecond timescale but becomes more relevant at femtosecond delays. Photoinduced Coulomb screening reduces the electron-hole interaction, leading to a decrease in exciton binding energy from approximately \SI{360}{\milli\electronvolt} at equilibrium to \SI{90}{\milli\electronvolt} at an excitation density of \SI{3.2e21}{\per\cubic\centi\meter}. This reduction manifests as a blue shift in the non-equilibrium absorption spectra (Fig. \ref{fig:CsPbBr3}a). These results highlight how the excitation density can be an effective parameter for tuning Coulomb screening and thereby enabling precise control over excitonic resonance energies.

\begin{figure*}
    \centering
    \includegraphics[width=0.8\linewidth]{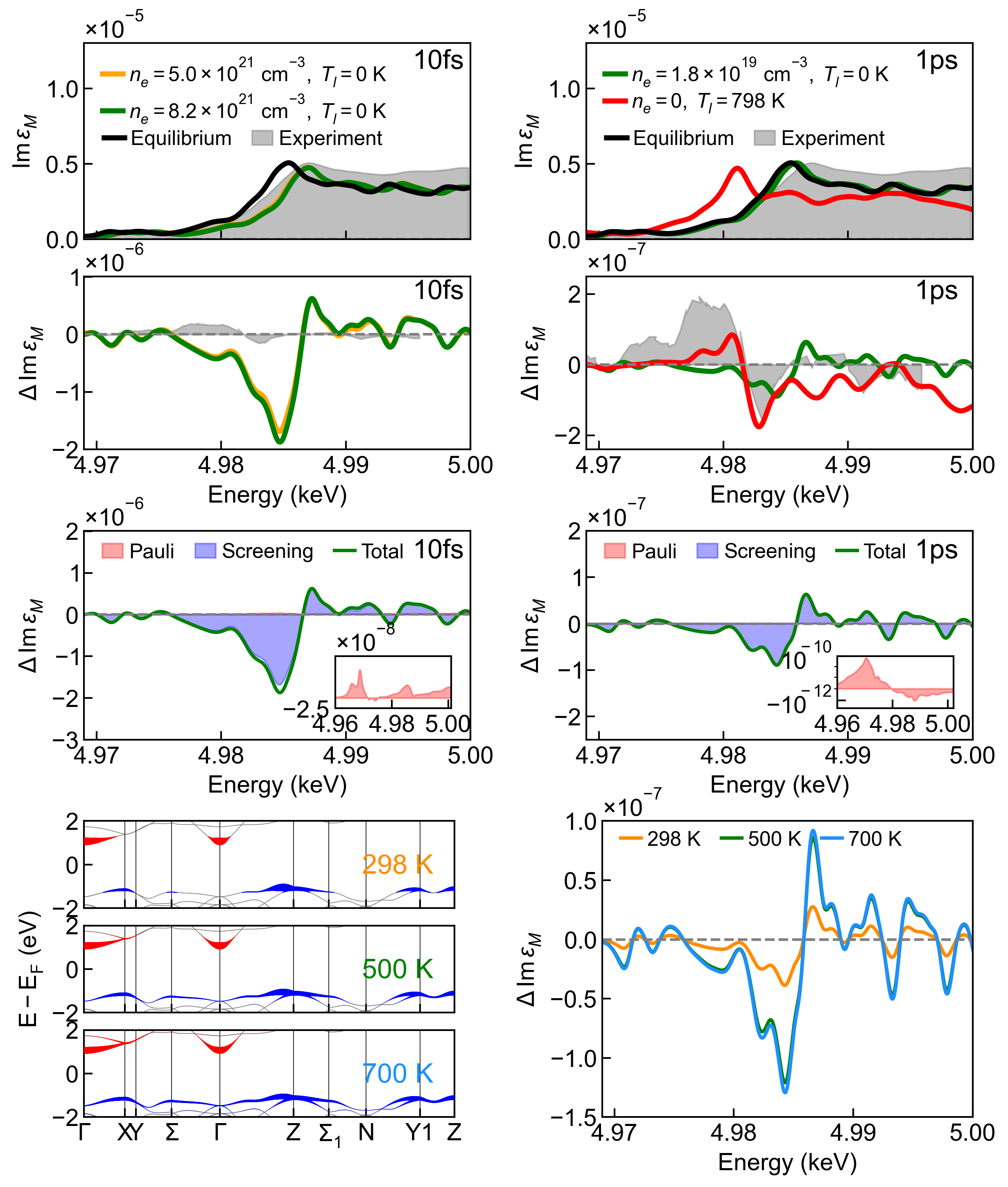}
    \caption{\textbf{Electronic and thermal contributions to TA spectra at the Ti K-edge of anatase TiO$_2$.} Same as Fig.~\ref{fig:WSe2}, but for the Ti K-edge of anatase TiO$_2$.}
    \label{fig:TiO2}
\end{figure*}

%%%%%%%%%%%%%%%%%%%%%%%%%%%%%%%%%%%%%%%%%%%%%%%%%%%%%%%%%%%%
%%%%%% Effect of carrier distribution effect on XTA %%%%%%%%
%%%%%%%%%%%%%%%%%%%%%%%%%%%%%%%%%%%%%%%%%%%%%%%%%%%%%%%%%%%%
We further investigate the effect of excited-carrier distribution on the Coulomb screening, using the Fermi-Dirac distribution at varying carrier temperatures (Fig.~\ref{fig:TiO2}e) for a fixed excitation density of \SI{8.0e16}{\per\cubic\centi\metre}. The resulting TA spectra are shown in Fig.~\ref{fig:TiO2}f. Higher carrier temperatures, corresponding to more delocalized distributions in $\mathbf{k}$-space, yield increased TA amplitudes that indicate enhanced Coulomb screening. This increase saturates around 500 K, similar to the case of WSe$_2$. This finding suggests that the carrier density and its distribution are both key factors in modulating photoinduced Coulomb screening and exciton resonance energies.\\

\noindent\textbf{Transient absorption at the Ti K-edge in anatase \texorpdfstring{TiO$_2$}{TiO2}.}
%%%%%%%%%%%%%%%%%%%%%%%%%%%%%%%%%%%%%%%%%%%%%%%%%%%%%%%%%%%%transient
%%%%%%%%%%%%%%%%% Introduction of TiO2 %%%%%%%%%%%%%%%%%%%%%
%%%%%%%%%%%%%%%%%%%%%%%%%%%%%%%%%%%%%%%%%%%%%%%%%%%%%%%%%%%%
Transition metal oxides, such as anatase titanium dioxide (TiO$_2$), are promising candidates for photocatalysis applications due to their strong light-matter interactions \cite{regan1991}. Understanding the ultrafast carrier dynamics is essential to improve its photocatalytic efficiency. X-ray TA spectroscopy has been widely used for this purpose \cite{rittmann2014,santomauro2015}, however, a complete theoretical interpretation of X-ray TA spectra is still missing. We provide such analysis, employing our newly developed approach to provide physical insights into the X-ray TA spectra of anatase TiO$_2$.

%%%%%%%%%%%%%%%%%%%%%%%%%%%%%%%%%%%%%%%%%%%%%%%%%%%%%%%%%%%%
%%%%%%%%%%%%%% absorption spectra at ps and fs timescale %%%%%%%%%%
%%%%%%%%%%%%%%%%%%%%%%%%%%%%%%%%%%%%%%%%%%%%%%%%%%%%%%%%%%%%
The equilibrium absorption spectra at the Ti K-edge are shown in the top panels of Figs. \ref{fig:TiO2}a-b, exhibiting good agreement with the experimental counterpart (gray-shaded area) \cite{santomauro2015}. At picosecond and femtosecond time delays, the spectra (orange and green lines) display a modest blue shift with increasing excitation density (top panels). At a lattice temperature of 798 K, significant red shifts of various features are observed, stemming from the reduced gap between the core-level and the conduction band. This results in a positive peak around \SI{4.982}{\kilo\electronvolt} in the TA spectra (bottom panels), a feature that cannot be accounted for by carrier excitation alone. The blue shift induced by carrier excitation yields a negative signal near \SI{4.984}{\kilo\electronvolt} (bottom panels of Figures \ref{fig:TiO2}a-b), which matches well with the negative peak in the experiment \cite{santomauro2015}.

%%%%%%%%%%%%%%%%%%%%%%%%%%%%%%%%%%%%%%%%%%%%%%%%%%%%%%%%%%%%%
%%%%%%%%%%%%%%%%   XTA spectra decomposition  %%%%%%%%%%%%%%%
%%%%%%%%%%%%%%%%%%%%%%%%%%%%%%%%%%%%%%%%%%%%%%%%%%%%%%%%%%%%%
Like for the other materials, we disentangle in Figs. \ref{fig:TiO2} c-d the two effects that shape the transient response. The photoinduced Coulomb screening (blue areas) dominates the TA spectra across both timescales, while Pauli blocking (red areas) plays a comparatively minor role. However, Pauli blocking is more noticeable on the femtosecond timescale, consistent with our observations in WSe$_2$ and CsPbBr$_3$. Due to Coulomb screening, the exciton binding energy decreases with increasing excitation density (Fig. \ref{figSI:TiO2_Eb}), blue-shifting the exciton resonance in the absorption spectra.

%%%%%%%%%%%%%%%%%%%%%%%%%%%%%%%%%%%%%%%%%%%%%%%%%%%%%%%%%%%%%
%%%%%%%%%%%% Effect of carrier distribution on XTA %%%%%%%%%%
%%%%%%%%%%%%%%%%%%%%%%%%%%%%%%%%%%%%%%%%%%%%%%%%%%%%%%%%%%%%%
Using the Fermi-Dirac statistics at different carrier temperatures for a fixed excitation density of \SI{1.8e19}{\per\cubic\centi\metre}, we further examine the influence of the excited carrier distributions on the Coulomb screening as illustrated in Fig.~\ref{fig:TiO2}e.  Figure~\ref{fig:TiO2}f presents the corresponding TA spectra. Higher carrier temperatures, corresponding to more delocalized distributions, lead to increased TA responses, indicative of enhanced Coulomb screening. Similar to WSe$_2$ and CsPbBr$_3$, a saturation is observed at 500 K.\\

\begin{figure*}
    \centering
    \includegraphics[width=1\linewidth]{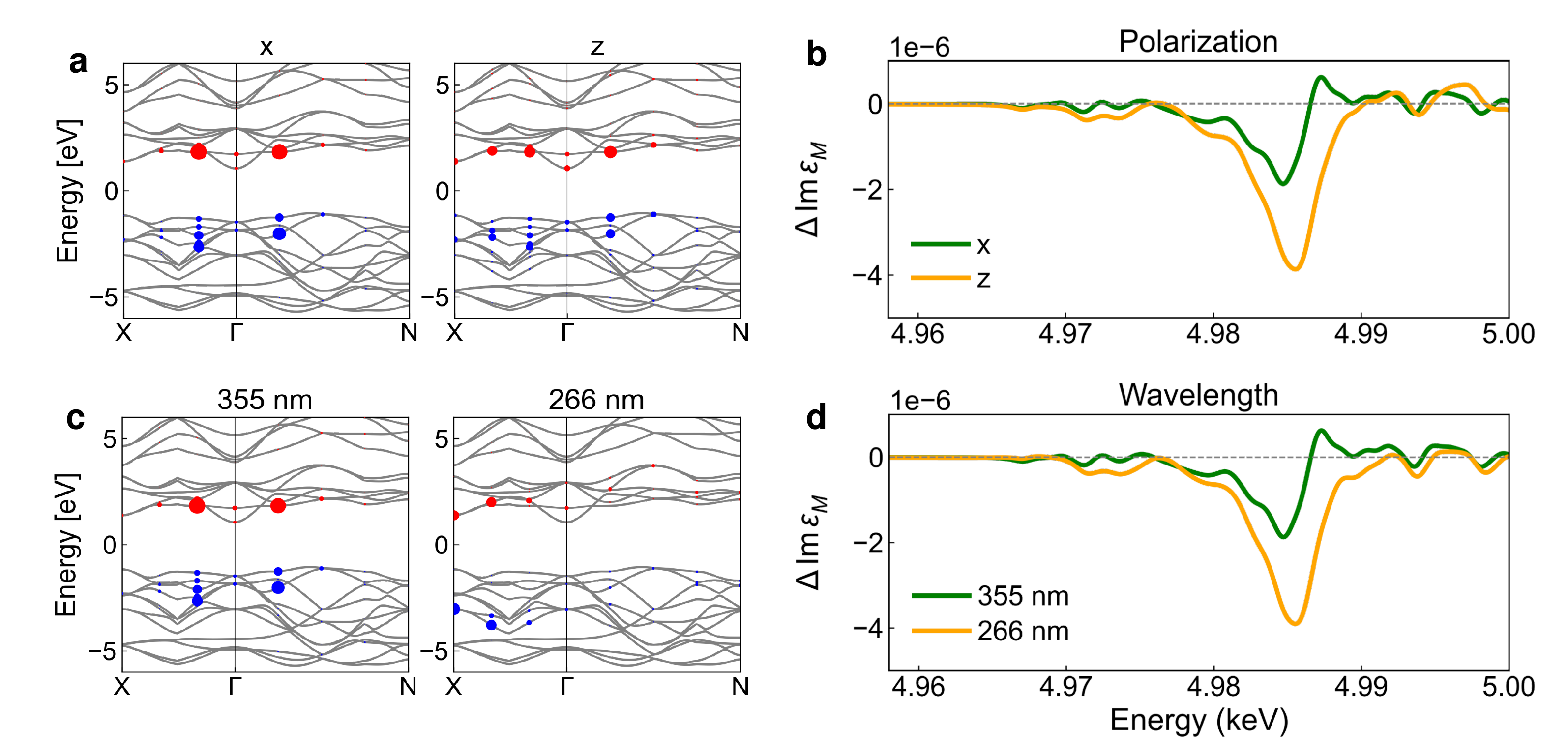}
    \caption{\textbf{Femtosecond control of core-exciton resonances by Coulomb screening.}  Time evolution of photoexcited electron (red) and hole (green) distributions following femtosecond optical excitation with varying (a) pump polarizations (along $x$ and $z$), and  (c) pump wavelengths (355~nm and 266~nm).  (d,f) Corresponding calculated TA spectra at the Ti K-edge in anatase TiO$_2$.}
    \label{fig:Tio2_polarization_wavelength}
\end{figure*}

\noindent\textbf{Control of photo-induced exciton screening.} 
For the three materials studied so far, we observed that higher excitation densities and more delocalized carrier distributions in $\mathbf{k}$-space enhance Coulomb screening by excited carriers. Beyond these two factors, we consider now the polarization direction of the pump beam. We take anatase TiO$_2$ as an example, as it exhibits a well-known anisotropic dielectric response, with stronger absorption when the electric field aligns with specific crystallographic directions \cite{kang2010}. At femtosecond delays, our simulations show that a pump polarization oriented along the x-direction yields a smaller excited carrier density (Fig. \ref{figSI:excitationDensity_time} a) compared to the polarization along z. In both cases, the degree of delocalization in $\mathbf{k}$-space is similar (Fig.~\ref{fig:Tio2_polarization_wavelength}a). The higher excitation density induced by a z-polarized pump gives rise to a stronger TA signal (panel b), indicating stronger Coulomb screening. This finding opens the possibility for tuning exciton resonances through the polarization of the pump beam.

The pump wavelength is another factor to explore, since it may also affect exciton resonances. Exciting anatase TiO$_2$ with different wavelengths at the same pump fluence results in nearly the same excitation densities (panel b), yet with a more pronounced TA spectrum for the shorter wavelength (panel d). The excitation with shorter wavelength populates the electronic states over a broader energy range, also further away from the band edges (panel c), which increases their polarizability and consequently enhances the dielectric screening. These findings identify the pump wavelength as an additional control parameter for tuning exciton resonance energies.

Controlling exciton screening in view of modulating exciton resonances offers broad opportunities for optoelectronic applications. Such control enables the tuning of the detection windows of X-ray detectors without altering the device structure. It also allows control over the energy range contributing to second-harmonic generation (SHG), thereby enabling the modulation of the generated second-harmonic wavelength. Overall, controlling exciton screening enables tunable light-matter interactions across a broad spectral range and is promising for designing photonic devices, such as photodetectors, nonlinear optical devices, and optical filters.\\

% where the sensitivity can be temporally tailored via external excitation.

In summary, we developed an approach that combines cDFT and RT-TDDFT with non-equilibrium BSE to simulate pump-probe spectroscopy and disentangle electronic and thermal contributions to the transient absorption spectra. We applied this method to three prototypical materials, WSe$_2$, CsPbBr$_3$, and anatase TiO$_2$, achieving excellent agreement with experiment. Our calculations reveal that thermal effects caused by local lattice heating induce red shifts in the exciton resonance energies. Photoinduced Coulomb screening is the dominant electronic effect that reduces exciton binding energies, resulting in blue shifts of the spectra. Finally, we demonstrated how the exciton resonance energies can be modulated on a femtosecond timescale by the choice of laser polarization and wavelength. All these parameters can be used for the design of energy-selective optoelectronic devices.  

The \emph{ab initio} formalism presented here can be applied to a wide range of semiconductors. It can capture ultrafast exciton dynamics across timescales ranging from attoseconds to picoseconds, spanning from the visible and near-infrared light to the extreme ultraviolet and the X-ray regime. To provide an outlook, our approach can also be applied to transient reflection, time-resolved terahertz spectroscopy, and potentially time-resolved photoluminescence, enabling comprehensive insight into the non-equilibrium exciton dynamics across diverse energy and time domains.\\

\clearpage
\noindent{\large\textbf{Methods}}\\
All calculations were performed using the all-electron, full-potential package \exciting, where the present non-equilibrium BSE formalism has been implemented. Ground-state electronic structures were obtained within the generalized gradient approximation using the Perdew–Burke–Ernzerhof (PBE) functional. Constrained DFT and RT-TDDFT were employed to generate thermal and non-thermal carrier distributions, respectively. The computed carrier distributions were subsequently used as input for the non-equilibrium BSE to obtain the non-equilibrium absorption spectra. Transient absorption (TA) spectra were then computed as the difference between these non-equilibrium spectra and the corresponding equilibrium reference, which defines the electronic contribution to the TA response. Thermal contributions were evaluated by computing absorption spectra for temperature-expanded lattices and subtracting the equilibrium spectrum. Additional computational details are summarized in the Supplementary Section~\ref{secSI:computational_details}.

\vspace{10pt}
\noindent{\large\textbf{Data availability}}\\
All input and output files are available in the NOMAD data infrastructure with the following link: https://doi.org/10.17172/NOMAD/2025.09.01-1

\vspace{10pt}
\noindent{\large\textbf{Acknowledgments}}\\  
Funding from the DFG, projects 182087777 and 424709454 is appreciated. LQ acknowledges funding from the Alexander von Humboldt Foundation. The authors gratefully acknowledge computing time on the high-performance computer ''Lise'' at the NHR center NHR@ZIB. This center is jointly supported by the Federal Ministry of Education and Research and the state governments participating in the NHR (\url{www.nhr-verein.de}).

\vspace{10pt}
\noindent{\large\textbf{Author contributions}}\\ 
L.Q., R.R.P, and C.D. conceptualized the work. R.R.P performed the calculations for WSe$_2$, L.Q. performed the calculations for TiO$_2$ and CsPbBr$_3$. R.R.P implemented the cDFT and RT-TDDFT methods. All authors contributed to the writing of the manuscript.

\vspace{10pt}
\noindent{\large\textbf{Competing interests}}\\ 
The authors declare no competing interests.

\vspace{10pt}
\noindent{\large\textbf{Additional information}}\\ 
\textbf{Supplementary information} The online version contains supplementary material available at

\vspace{10pt}
\noindent\textbf{Correspondence} Correspondence and requests should be addressed to Claudia Draxl.

\bibliography{references-NC}% Produces the bibliography via BibTeX.

\newpage
\newpage
\clearpage
{
\onecolumngrid
\center
\Huge{First-principles approach to ultrafast pump-probe spectroscopy in solids \\ Supporting Information}

%\affil[1]{Department of Physics and CSMB, Humboldt-Universit\"at zu Berlin, Zum Gro\ss en Windkanal 2, D-12489 Berlin, Germany}
%\affil[2]{Supercomputing Department, Zuse Institute Berlin (ZIB), Takustraße 7, 14195 Berlin, Germany}
}
\setcounter{equation}{0}
\renewcommand{\theequation}{S.\arabic{equation}}
\renewcommand{\thefigure}{S.\arabic{figure}}
\section{Theory}
\label{secSI:theory_background}
\subsection{Constrained density functional theory}

\noindent \hspace{1em} The Kohn-Sham (KS) equations of density-functional theory (DFT) provide a practical method for determining auxiliary single-particle wavefunctions, $\psi_{i\mathbf{k}}(\mathbf{r})$, and energies $\epsilon_{i\mathbf{k}}$:
\begin{equation}
	\left[-\frac{\nabla^2}{2}+v_{\text{KS}}(\mathbf{r})\right]\psi_{i\mathbf{k}}(\mathbf{r})=\epsilon_{i\mathbf{k}}\psi_{i\mathbf{k}}(\mathbf{r}),
\end{equation}
where $\mathbf{k}$ denotes the wavevector, and $v_{\text{KS}}(\mathbf{r})$, the KS potential. In this framework, the ground-state electron density $n_{0}(\mathbf{r})$ is be expressed as:
\begin{equation}
	n_{0}(\mathbf{r})=\sum_\mathbf{k}w_\mathbf{k}\sum_if_{i\mathbf{k}}^0|\psi_{i\mathbf{k}}(\mathbf{r})|^2,
\end{equation}
where $w_\mathbf{k}$ is the $\mathbf{k}$-point weight, and $f_{i\mathbf{k}}^0$ is the occupation number of the KS state $i\mathbf{k}$. 

In constrained density-functional theory (cDFT), excited states are modeled by considering that electrons have been promoted to empty states, leaving holes in the previously occupied states. The resulting excited-state electron density $n(\mathbf{r})$ is given by:
\begin{equation}
	n(\mathbf{r})=\sum_\mathbf{k}w_\mathbf{k}\left(\sum_if_{i\mathbf{k}}^0|\psi_{i\mathbf{k}}(\mathbf{r})|^2+\sum_{c}f_{c\mathbf{k}}|\psi_{c\mathbf{k}}(\mathbf{r})|^2-\sum_{v}f_{v\mathbf{k}}|\psi_{v\mathbf{k}}(\mathbf{r})|^2
	\right),
\end{equation}
where $v$ and $c$ represent the indices of valence and conduction states, respectively. The distribution of excited electrons ($f_{c\mathbf{k}}$) and holes ($f_{v\mathbf{k}}$) is constrained based on some physical insight that fulfills the following condition:
\begin{equation}\label{eq:cdft_nexc}
	\sum_{c\mathbf{k}}w_\mathbf{k}f_{c\mathbf{k}} = \sum_{v\mathbf{k}}w_\mathbf{k}f_{v\mathbf{k}} = N_\mathrm{exc},
\end{equation}
where $N_\mathrm{exc}$ stands for the number of excitations per unit cell.

%%%%%%%%%%%%%%%%%%%%%%%%%%%%%%%%%%
% RT-TDDFT
%%%%%%%%%%%%%%%%%%%%%%%%%%%%%%%%%%
\subsection{Real-time time-dependent density-functional theory}

\noindent \hspace{1em} In real-time time-dependent density-functional theory (RT-TDDFT), the time evolution of electrons subjected to time-dependent perturbations, such as electric or magnetic fields, is given by:
\begin{equation}
	\frac\partial{\partial t}\psi_{j\mathbf{k}}(\mathbf{r},t) = -\mathrm{i}\hat{H}(\mathbf{r},t)\psi_{j\mathbf{k}}(\mathbf{r},t),
\end{equation}
where $\hat{H}(\mathbf{r},t)$ and $\psi_{j\mathbf{k}}(\mathbf{r},t)$ are the time-dependent KS Hamiltonian and wavefunction, respectively. Considering an external perturbation as a laser pulse, with the electric field described by the vector potential $\mathbf{A}(t)$, the corresponding KS Hamiltonian can be expressed as \cite{Yabana2012:19771}:
\begin{equation}
	\hat{H}(\mathbf{r},t)=\frac12\left(-\mathrm{i}\nabla+\frac1c\mathbf{A}(t)\right)^2+v_{\text{KS}}(\mathbf{r},t), 
\end{equation}
\noindent with $v_{\text{KS}}(\mathbf{r},t)$ being the time-dependent KS potential.

Monitoring the time-resolved creation of electron-hole pairs over time is essential for comprehending excitation dynamics. In RT-TDDFT, this can be obtained, similar to Eq. (\ref{eq:cdft_nexc}), from time-dependent distributions, $f_{c\mathbf{k}}(t)$ and $f_{v\mathbf{k}}(t)$. These are determined by projecting $\psi_{i\mathbf{k}}(\mathbf{r},t)$ onto the KS states at the initial time, $\psi_{i\mathbf{k}}(\mathbf{r}, 0)$ \cite{RodriguesPela2021:86963}: 
\begin{equation}
	f_{c\mathbf{k}}(t)=\sum_i f^0_{i\mathbf{k}}|\langle\psi_{c\mathbf{k}}(0)|\psi_{i\mathbf{k}}(t)\rangle|^{2} ,
\end{equation}
\begin{equation}
	f_{v\mathbf{k}}(t)=f_{v\mathbf{k}}^0-\sum_{i\mathbf{k}} f^0_{i\mathbf{k}} |\langle\psi_{v\mathbf{k}}(0)|\psi_{i\mathbf{k}}(t)\rangle|^2.
\end{equation}

%%%%%%%%%%%%%%%%%%%%%%%%%%%%%%%%%%%
% BSE
%%%%%%%%%%%%%%%%%%%%%%%%%%%%%%%%%%% 

\subsection{Bethe-Salpeter equation}

\noindent \hspace{1em} Absorption spectra are computed by solving the Bethe-Salpeter equation (BSE) for the two-particle Green's function. In matrix form, the BSE takes the form of an eigenvalue equation \cite{Rohlfing_2000}:
\begin{equation}  \sum_{o^{\prime}u^{\prime}\mathbf{k}^{\prime}}H_{ou\mathbf{k},o^{\prime}u^{\prime}\mathbf{k}^{\prime}}^{BSE}A_{o^{\prime}u^{\prime}\mathbf{k}^{\prime}}^{\lambda}=E_{\lambda}A_{ou\mathbf{k}}^{\lambda},
	\label{Schrödinger equation}
\end{equation}
where $E_{\lambda}$ denotes the excitation energy and $A^{\lambda}_{ou\mathbf{k}}$ are corresponding eigenvectors, indices $o$ and $u$ label the occupied and unoccupied states, respectively. 

The $H^{BSE}$ Hamiltonian is written as:
\begin{equation}
	H^{BSE}=H^{diag}+2H^{x}+H^{dir}.
\end{equation}
$H^{diag}$ is the diagonal term representing independent-particle transitions; $H^{x}$ is the exchange term including the repulsive bare Coulomb interaction; and $H^{dir}$ is the direct term containing the attractive screened Coulomb interaction:
\begin{equation}    H_{ou\mathbf{k},o^{\prime}u^{\prime}\mathbf{k^{\prime}}}^{diag}=\left(\epsilon_{u\mathbf{k}}-\epsilon_{o\mathbf{k}}\right)\delta_{oo^{\prime}}\delta_{uu^{\prime}}\delta_{\mathbf{kk^{\prime}}},
\end{equation}

\begin{equation}
	H_{vo\mathbf{k},v^{\prime}o^{\prime}\mathbf{k}^{\prime}}^{x}  \\  =\int d^3\mathbf{r}d^3\mathbf{r^\prime}\psi_{o\mathbf{k}}(\mathbf{r})\psi_{u\mathbf{k}}^{*}(\mathbf{r})\nu(\mathbf{r},\mathbf{r}^{\prime})\psi_{o^{\prime}\mathbf{k}^{\prime}}^{*}(\mathbf{r}^{\prime})\psi_{u^{\prime}\mathbf{k}^{\prime}}(\mathbf{r}^{\prime}),
\end{equation}

\begin{equation}
	H_{ou\mathbf{k},o^{\prime}u^{\prime}\mathbf{k^{\prime}}}^{dir}  =-\int d^3\mathbf{r}d^3\mathbf{r^\prime}\psi_{o\mathbf{k}}(\mathbf{r})\psi_{u\mathbf{k}}^{*}(\mathbf{r^{\prime}})W(\mathbf{r},\mathbf{r^{\prime}})\psi_{o^{\prime}\mathbf{k^{\prime}}}^{*}(\mathbf{r})\psi_{u^{\prime}\mathbf{k^{\prime}}}(\mathbf{r^{\prime}}).
	\label{H_direct_term}
\end{equation}

\noindent $\nu(\mathbf{r},\mathbf{r}^{\prime})$ and $W(\mathbf{r},\mathbf{r}^{\prime})$ stand for the bare and screened Coulomb potentials, respectively. 

\paragraph{Electronic contribution:} To capture non-equilibrium effects in the spectra, carrier occupations obtained from cDFT and RT-TDDFT are explicitly included. By selectively incorporating these occupations either in the screened Coulomb interaction or in the dipole matrix elements, we can isolate the purely electronic contribution (Coulomb screening and Pauli blocking) to the transient spectra.

The screened Coulomb potential in reciprocal space is given by:
\begin{equation}
	W_{\mathbf{GG}^{\prime}}(\mathbf{q})=4\pi\frac{\varepsilon_{\mathbf{GG}^{\prime}}^{-1}(\mathbf{q},\omega=0)}{|\mathbf{q}+\mathbf{G}||\mathbf{q}+\mathbf{G^{\prime}}|},
\end{equation}
where the microscopic dielectric matrix $\varepsilon_{\mathbf{GG}'}$, is evaluated within the random-phase approximation (RPA) as
\begin{equation}
	\varepsilon_{\mathbf{GG}^{\prime}}(\mathbf{q},\omega)=\delta_{\mathbf{GG}^{\prime}}-\frac1V\nu_{\mathbf{G}^{\prime}}(\mathbf{q})\sum_{ou\mathbf{k}}\frac{f_{u\mathbf{k}+\mathbf{q}}-f_{o\mathbf{k}}}{\epsilon_{u\mathbf{k}+\mathbf{q}}-\epsilon_{o\mathbf{k}}-\omega}\left[M_{ou}^{\mathbf{G}}(\mathbf{k},\mathbf{q})\right]^*M_{ou}^{\mathbf{G}^{\prime}}(\mathbf{k},\mathbf{q}).
	\label{eq:CoulombScreening}
\end{equation}

\noindent where $V$ denotes the unit cell volume, $f_{o\mathbf{k}}$ represents the excited-state occupation, and $M_{ou}^{\mathbf{G}}(\mathbf{k},\mathbf{q})$ is the plane-wave matrix element \cite{Vorwerk:2017gs, Vorwerk_2019}.

Excited-state occupations also enter the dipole matrix elements, ($\mathrm{D}^*$) , defined as:
\begin{equation}
	\mathrm{D}^* =\mathrm{i}\sqrt{|f_{o\mathbf{k}}-f_{u\mathbf{k}+\mathbf{q}}|} \times \frac{\langle c\mathbf{k}|\hat{\mathbf{p}}|u\mathbf{k}\rangle}{\epsilon_{u\mathbf{k}+\mathbf{q}} - \epsilon_{o\mathbf{k}}},
	\label{eq: PauliBlocking}
\end{equation}
leading to the transition coefficient, $t_\lambda$:
\begin{equation}
	t_\lambda(0,\mathbf{q})=-\mathrm{i}\frac{\hat{\mathbf{q}}}{|\mathbf{q}|}\mathbf{X}_\lambda^\dagger\mathrm{D}^*,
\end{equation}
where $\mathbf{X}_\lambda$ is the resonant part of eigenvectors. The $\beta\beta$ tensor components of the macroscopic dielectric tensor is:
\begin{equation}
	\begin{aligned}
		\varepsilon_M^{\beta\beta}(\omega)& =1-\frac{8\pi}{V}\sum_\lambda|t_{\lambda}^\beta|^2\delta(\omega-E_\lambda).
	\end{aligned}
\end{equation}

%\section{Implementation}

%%%%%%%%%%%%%%%%%%%%%%%%%%%%%%%%%%%%%%%%%%%%%%%%%%%%%%%%
%%%%%%%%%%%%%%%%%%%%%%%%%%%%%%%%%%%%%%%%%%%%%%%%%%%%%%%%
\section{Computational details}
%%%%%%%%%%%%%%%%%%%%%%%%%%%%%%%%%%%%%%%%%%%%%%%%%%%%%%%%
%%%%%%%%%%%%%%%%%%%%%%%%%%%%%%%%%%%%%%%%%%%%%%%%%%%%%%%%
\label{secSI:computational_details}

\subsection{Ground-state calculation}
Brillouin zone integrations were performed using a Monkhorst-Pack \textbf{k}-point grid of $18 \times 18 \times 5$ for WSe$_2$, $8 \times 8 \times 6$ for TiO$_2$ and $6 \times 6 \times 4$ for CsPbBr$_3$. The Perdew–Burke–Ernzerhof (PBE) functional within the generalized gradient approximation (GGA) was employed to describe exchange-correlation effects. Scalar relativistic effects, including the mass-velocity and Darwin terms, were treated using the atomic ZORA (zero-order regular approximation) scheme, while spin-orbit coupling was neglected. 

\subsection{Excited-state calculation}
\begin{table}[H]
	\centering
	\caption{Computational parameters used in cDFT, RT-TDDFT, and BSE calculations for WSe$_2$, CsPbBr$_3$, and TiO$_2$.}
	\label{tab:comp_params}
	\begin{tabular}{llccc}
		\toprule
		\textbf{Method} & \textbf{Parameter} & \textbf{WSe$_2$} & \textbf{CsPbBr$_3$} & \textbf{TiO$_2$} \\
		\midrule
		\multirow{2}{*}{cDFT} 
		& $k$-point mesh            & $9\times9\times2$    & $6\times6\times4$     & $8\times8\times6$ \\
		& $R_{\mathrm{MT}}|\mathbf{G}+\mathbf{k}|_\mathrm{max}$   & 8.0 & 8.0  & 8.0 \\
		\midrule
		\multirow{5}{*}{RT-TDDFT} 
		& $k$-point mesh            & $9\times9\times2$    & $6\times6\times4$     & $8\times8\times6$ \\
		& $R_{\mathrm{MT}}|\mathbf{G}+\mathbf{k}|_\mathrm{max}$   & 8.0 & 8.0  & 8.0 \\
		& Photon energy (eV)        & 1.9    & 3.49     & 3.49  \\
		& Polarization direction    & z    & x       & x      \\
		& Pulse duration (fs)       & 4    & 1       & 10     \\
		\midrule
		\multirow{5}{*}{BSE} 
		& $k$-point mesh            & $9\times9\times2$    & $6\times6\times4$     & $8\times8\times6$ \\
		& Local-field cutoff $|\mathbf{G}+\mathbf{q}|_\mathrm{max}$ (a.u.$^{-1}$) 
		& 2.5    & 2.5     &4.0   \\
		& broadening (eV)          & 0.2    & 2.4   & 0.9  \\
		& Number of empty states    & 100    & 328     & 62    \\
		& Scissor shift (eV) - 0 K        & 6.9    & 163.5   & 106.2   \\
		& Scissor shift (eV) - high temperature        & 6.5    & 162.0   &  104.5  \\
		\bottomrule
	\end{tabular}
\end{table}

%%%%%%%%%%%%%%%%%%%%%%%%%%%%%%%%%%%%%%%%%%%%%%%%%%%%%%%%
%%%%%%%%%%%%%%%%%%%%%%%%%%%%%%%%%%%%%%%%%%%%%%%%%%%%%%%%
\newpage
\section{Additional results}
\label{secSI:Computational results}
\subsection{TiO$_2$}
%%%%%%%%%%%%%%%%%%%%%%%%%%%%%%%%%%%%%%%%%%%%%%%%%%%%%%%%
%%%%%%%%%%%%%%%%%%%%%%%%%%%%%%%%%%%%%%%%%%%%%%%%%%%%%%%%

\begin{figure}[H]
	\centering
	\includegraphics[width=0.5\linewidth]{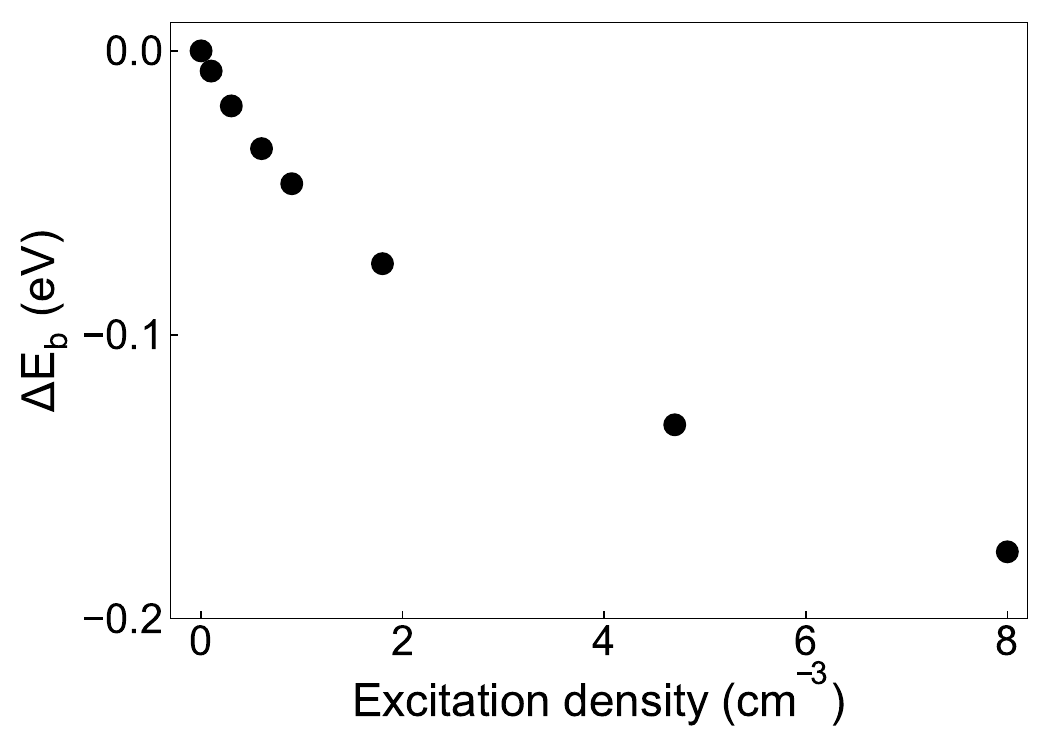}
	\caption{Change in Ti 1s core-exciton binding energy in TiO$_{2}$ as a function of excitation density, relative to the equilibrium binding energy.}
	\label{figSI:TiO2_Eb}
\end{figure}

\begin{figure}[H]
	\centering    
	\includegraphics[width=0.49\linewidth]{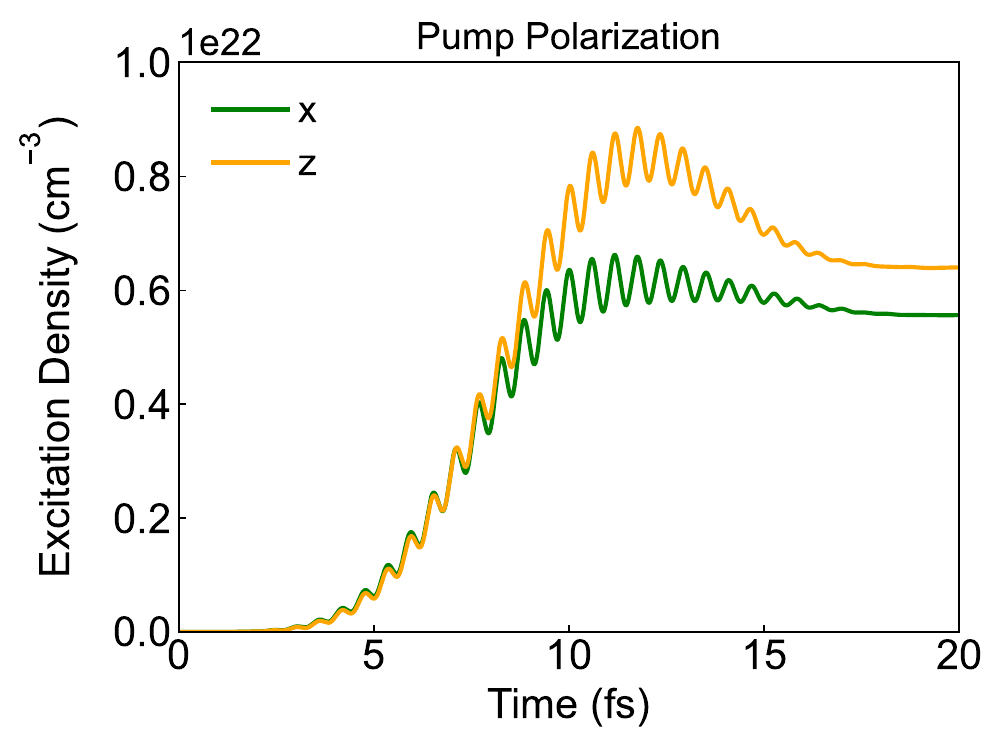}    \includegraphics[width=0.49\linewidth]{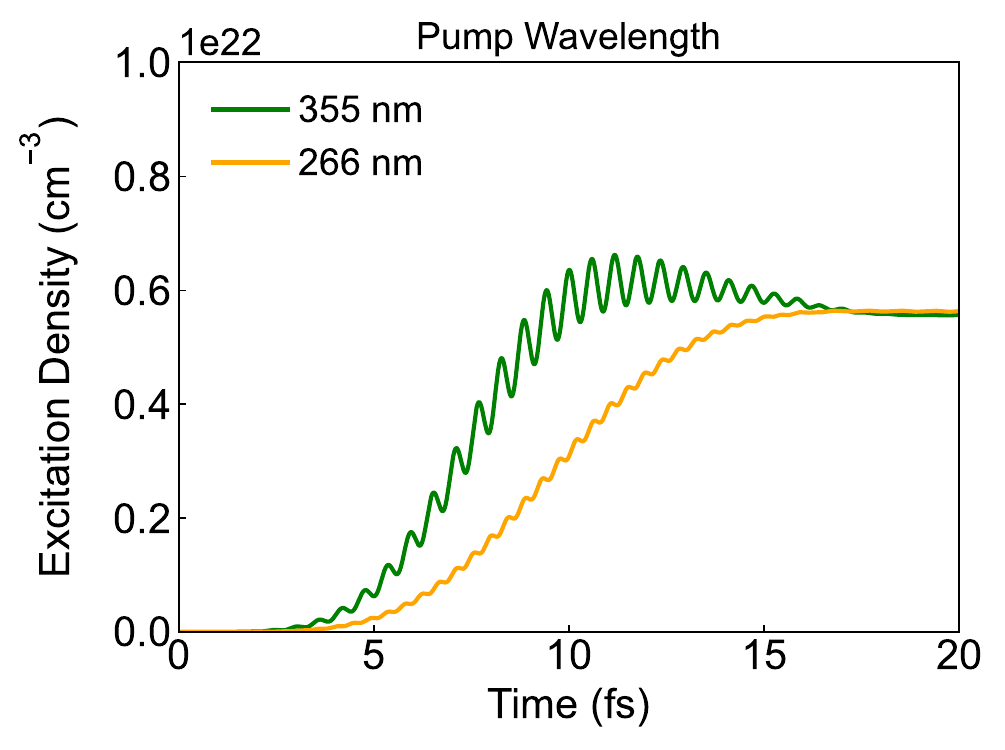}    
	\caption{Temporal evolution of the excitation density following pump excitation under different (a) pump polarization directions and (b) excitation photon energies.}
	\label{figSI:excitationDensity_time}
\end{figure}

\clearpage

\subsection{WSe$_2$}

\begin{figure}[H]
	\centering   
	\includegraphics[scale=1]{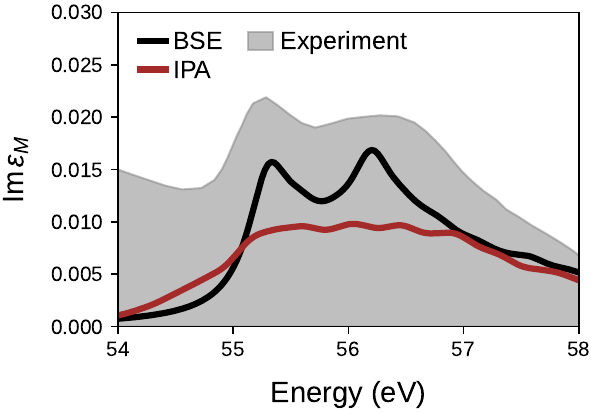}    
	\caption{Imaginary part of the dielectric function of WSe$_2$ at the $M_{4,5}$ edges, compared with experiment \cite{Oh_2023}. Two theoretical approaches are shown: the BSE, which accounts for electron-hole interaction, and the independent-particle approximation (IPA).}
	\label{figSI:xas_WSe2_BSE_vs_IPA}
\end{figure}

%\bibliographystyle{apsrev4-2}
%\bibliography{references-NC}

\end{document}